\documentstyle[twocolumn,prl,aps,epsfig,amssymb]{revtex}
\tighten
\let\jnfont=\rm
\def\NPB#1,{{\jnfont Nucl.\ Phys.\ B }{\bf #1},}
\def\PLB#1,{{\jnfont Phys.\ Lett.\ B }{\bf #1},}
\def\EPJC#1,{{\jnfont Eur.\ Phys.\ J.\ C }{\bf #1},}
\def\PRD#1,{{\jnfont Phys.\ Rev.\ D }{\bf #1},}
\def\PRL#1,{{\jnfont Phys.\ Rev.\ Lett.\ }{\bf #1},}
\def\MPLA#1,{{\jnfont Mod.\ Phys.\ Lett.\ A }{\bf #1},}
\def\JPG#1,{{\jnfont J.\ Phys.\ G}{\bf #1},}
\def\CTP#1,{{\jnfont Commun.\ Theor.\ Phys.\ }{\bf #1},}

\begin{document}
\draft
\preprint{}

\title{Gravitino dark matter from gluino late decay in split supersymmetry}                                                                                
                                                                                
\author{Fei Wang $^1$,  Wenyu Wang $^1$,  Jin Min Yang $^{2,1}$}
                                                                                
\address{ \ \\[2mm]
{\it $^1$ Institute of Theoretical Physics, Academia Sinica, Beijing 100080, China} \\ [2mm]
{\it $^2$ CCAST(World Laboratory), P.O.Box 8730, Beijing 100080, China} \ \\[6mm] }
 
\date{\today}
 
\maketitle

\begin{abstract}
In split-supersymmetry (split-SUSY), gluino is a metastable particle and thus
can freeze out in the early universe. The late decay of such a long-life gluino
into the lightest supersymmetric particle (LSP) may provide much of the cosmic dark 
matter content. 
In this work, assuming the LSP is gravitino produced from the late decay of the 
metastable gluino, we examine the WMAP dark matter constraints on the gluino mass.
We find that to provide the full abundance of dark matter, the gluino must be heavier 
than about 14 TeV and thus not accessible at the CERN Large Hadron Collider (LHC). 
\end{abstract}


\section{Introduction}
In the recently proposed split-SUSY \cite{split}, inspired by the need of fine-tuning for 
the cosmological constant, the authors argued that the fine-tuning problem in particle 
physics does not have to be solved by SUSY. The only phenomenological constraints on split-SUSY
are then from the grand unification consideration as well as the dark matter consideration.
As a result, the sfermion mass scale can be very high while the gaugino/Higgsino 
mass scale is still around the weak scale. While the split-SUSY has the obvious 
virtue of naturally solving the notorious SUSY flavor problem,  it predicts that no sfermions 
are accessible at the CERN LHC collider. Thus if split-SUSY is indeed chosen by nature,
the only way to reveal SUSY at the LHC is through gaugino or Higgsino productions, especially
the gluino production \cite{gluino}.
This makes it important to pre-examine the possible mass range of these particles 
before the running of the LHC.

It is interesting that although the gauginos and Higgsinos in split-SUSY are required to be 
relatively light, they are recently found not necessarily below TeV scale 
from the grand unification and dark matter requirements \cite{senatore}.
Actually, the grand unification requirement can allow a heavy gaugino mass as high as 
18 TeV \cite{senatore}. If all gauginos and Higgsinos are above TeV scale, the LHC is 
doomed to find no SUSY particles except a light Higgs boson if split-SUSY is true.
Although the split-SUSY consequence in the dark matter issue is also considered in 
the literature \cite{senatore,pierce,profumo}, the authors focused on neutralino LSP 
or the usual NLSP decaying to the LSP during the BBN era, which    
is severely constrained by the Big Bang Nucleosythesis (BBN) and Cosmic Microwave Background 
(CMB) in split-SUSY. In this work we study the
dark-matter consequence of the long-life gluino in split-SUSY.  

In the usual weak-scale SUSY, the LSP is usually assumed to be the 
lightest neutralino \footnote{If gluino is the LSP, its relic abundance 
is severely constrained by the bounds from existing anomalous heavy isotope abundances 
\cite{baer,raby}.}. Gluino decays rapidly into the LSP and thus cannot freeze out to cause 
any dark matter consequence. 
Only in case that gluino is quasi-degenerate with the neutralino LSP can it have dark matter 
consequence through gluino-neutralino co-annihilation \cite{coannihilation}.

In split-SUSY, however, due to its long lifetime, gluino can freeze out before decaying and 
then decay slowly into the LSP, providing much of the cosmic dark matter content. 
So the gluino late decay is one characteristic of split SUSY.
In this work, assuming the LSP is the gravitino (the so-called superWIMP dark matter)
produced from the late decay of the metastable gluino, we will examine the  
WMAP dark matter constraints on the gluino mass.

Note that if the gluino lifetime is too long (as long as BBN time), the released energy 
from its decay may spoil the BBN success and also affect CMB as well as large scale structure 
formation \cite{arvanitaki}. Therefore, in our study we require that the gluino decays before 
BBN. 

\section{Calculations}
The gluino relic density from thermal production can be calculated from the Boltzmann equation 
   \begin{equation} 
   \frac{d n}{d t}= -3 H n -<\sigma v> (n^2-n_{eq}^2) , 
   \end{equation} 
where $H$ is the Hubble constant, $n$ is the particle number density of gluino,
$n_{eq}$ is the equilibrium density, and $<\sigma v>$ is the thermal averaged cross section 
of gluino annihilation. We can employ the freeze-out approximation technique to calculate 
the relic abundance.

In split-SUSY the gluino pair annihilation proceeds through 
the $s$-channel gluon-exchange diagram and the $t$-channel gluino-exchange 
diagram, as shown in Fig.~1(a-c). The squark-exchange diagrams shown in Fig.~1(d)
drop out since they are suppressed by the superheavy squark masses.
\begin{figure}[tb]
\begin{center} \epsfig{file=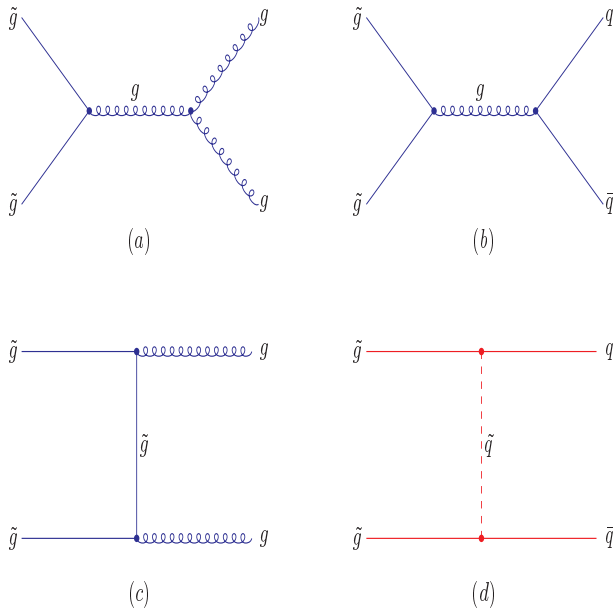,width=9.0cm, height=8.7cm}\end{center}
\caption{Typical Feynman diagrams of gluino pair annihilation into the SM particles.
         The last diagram through exchanging a squark makes negligible contribution
         in split-SUSY due to the superheavy squarks.}
\end{figure}
The perturbation annihilation cross-section reads \cite{baer} 
\begin{eqnarray} \label{width1}
\sigma\left(\tilde{g}\tilde{g}\to g g \right) & = &
  \frac{3 \pi \alpha_{s}^2}{16 {\beta}^2 s} \left\{ \log\frac{1+\beta}{1-\beta}
  \left[ 21-6{\beta}^2-3 {\beta}^4 \right] \right. \nonumber \\ 
&& \left. -33\beta+17 {\beta}^3 \right\} ,   \\
 \label{width2}
\sigma\left(\tilde{g}\tilde{g}\to q \bar{q} \right)& = & 
  \frac{ \pi \alpha_{s}^2 \bar{\beta} }{16 {\beta} s} (3-{\beta}^2)(3-{\bar{\beta}}^2) ,
\end{eqnarray}
where $\beta=\sqrt{1-4m_{\tilde{g}}^2/s}$ and $\bar{\beta}=\sqrt{1-4m_{q}^2/s}$
with $m_{\tilde{g}}$ and $m_{q}$ being the gluino mass and quark mass, respectively.
Multiple gluon exchanges between interacting $\tilde{g}$ will give rise to a Sommerfeld 
enhancement factor \cite{appelquist}
\begin{equation} 
 E=\frac{ C \pi \alpha_{s} }{\beta}
    {\left[1-\exp\left\{- \frac{ C \pi \alpha_{s} }{\beta}\right\} \right]}^{-1} .
\end{equation}
The interaction between the gluino and the goldstino (spin 1/2 component of gravitino) is 
suppressed by $1/F$ where $F$ characterizes the SUSY breaking scale. 
The gluino decay width to goldstino in split-SUSY is given by \cite{gambino}
\begin{equation} 
 \Gamma_{\tilde{G}g}=\frac{m_{\tilde{g}}^3}{2 \pi F^2} {C_{5}^{\tilde{G}}}^2 ,
\end{equation}
where $C_{5}^{\tilde{G}}=-m_{\tilde g}/2 \sqrt{2}$. Since the decay width is suppressed by
$1/F$, not necessarily $1/M_{pl}$, the gluino decay can be arranged to occur before BBN 
by choosing the value of $F$. Note the conventional superWIMP dark matter scenario 
(gravitino is the LSP) is severely constrained by BBN and CMB \cite{feng} since the 
late decay of NLSP to LSP is assumed to occur at $10^6\sim 10^8$ second and the released 
energy may spoil the success of standard BBN.
In our study we avoided such severe constraints since we require the gluino decays 
before BBN time. Furthermore, we also require the gluino decays before QCD era.
Otherwise, R-hadrons could be formed and the R-hadron annihilation could destroy 
gluinos \cite{arvanitaki}. 

The relic density of the gravitino LSP from the late decay of gluino is 
given by  
 \begin{equation} 
 \Omega_{\tilde{g}}\frac{m_{\tilde{G}}}{m_{\tilde{g}}} ,
 \end{equation}
 with the gravitino mass given by
 \begin{equation} 
  m_{\tilde{G}}=\sqrt{\frac{8 \pi}{3}}\frac{F}{M_{pl}}. 
 \end{equation}
 
Note that the gravitino can also be thermally produced \cite{relic} at the very early universe
with temperature $T\sim M_{pl}$ (or even a bit lower) and then freeze out when temperature drops.
But between the time of gravitino generation and now, the universe is expected to experience 
an inflation. Such an inflation would dilute the thermal relic density of gravitino.
So we neglect the gravitino thermal production at the very early universe.  
In the context of inflation, the universe is expected to be reheated after inflation.  
The gravitino can be generated from reheating if the reheating temperature is
high enough \cite{reheat}. 
In our study, we ignore such gravitino production by assuming the reheating 
temperature is not high enough to generate gravitino.

The cosmic non-baryonic dark matter relic density can be obtained from the
Wilkinson Microwave Anisotropy Probe (WMAP) measurements \cite{wmap} 
\begin{eqnarray}
       \Omega_m  =  0.27_{-0.04}^{+0.04} \ , ~~
       \Omega_b =  0.044_{-0.004}^{+0.004} \ ,
\end{eqnarray}
where  $\Omega_m$ is the total matter density and $\Omega_b$ is the baryonic matter density.
Requiring the gravitino dark matter abundance from the gluino late decay is within the  
$2\sigma$ range of the WMAP data, we obtain the allowed parameter space  
in the plane of $M_{LSP}$ versus $M_{\tilde g}$, as shown in Fig. 2.

We see from Fig. 2 that if the gravitino LSP from the gluino late-decay is to 
account for the whole gravitino dark matter content, the gluino has to be heavier  
than about 14 TeV and gravitino has to be lighter than about 16 TeV.

A few remarks are in order regarding the above results.
\begin{figure}[tb] 
\hspace*{-1.0cm} \epsfig{file=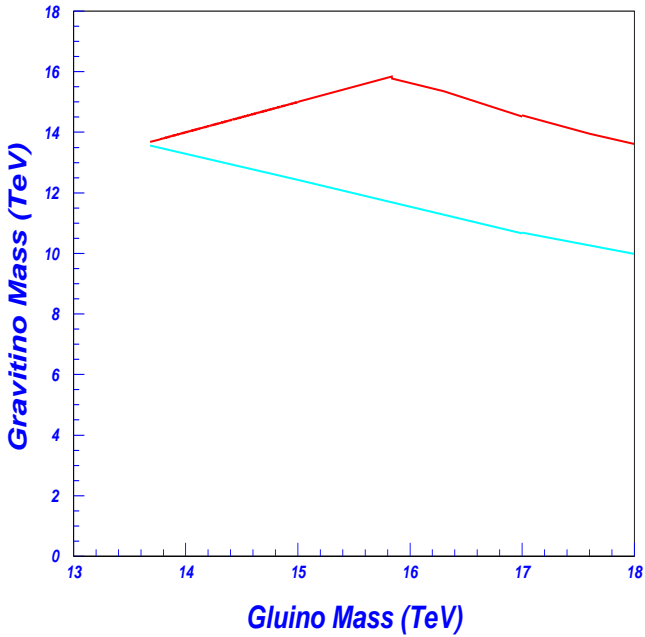,width=9.0cm, height=8.7cm}
\vspace*{0.3cm}
\caption{ The region between the two curves is the $2\sigma$ range  allowed by 
          the WMAP dark matter data.}
\end{figure}
(1) In our analyses the gluino is essentially assumed to be the next-to-lightest 
           supersymmetric particle (NLSP). If it is not the NLSP, it would 
           decay dominantly to the NLSP (say the neutralino $\tilde \chi^0_1$)
           followed by the decay of the NLSP to the gravitino LSP. In this
           case, although the gluino late decay also contributes to the dark
           matter content, its contribution is small compared to the freeze-out
           of neutralino NLSP.                 

(2) Although the gluino is assumed to be the NLSP, the
           lightest neutralino $\tilde \chi^0_1$ (assumed to be heavier than 
           gluino) can still freeze out since its decay to gluino is suppressed by
           heavy squark mass in split-SUSY. Of course, the  neutralino freeze-out
           happens much earlier than gluino freeze-out since its interaction is
           much weaker. Depending on the lifetime of the neutralino, its
           dark matter consequence can be quite different. If its decay to gluino 
           happens before the freeze-out of gluino (corresponding to the relatively 
           light squark mass), then the relic density of gluino is from the freeze-out,
           as assumed in our analyses. If its decay to gluino 
           happens after the freeze-out of gluino (corresponding to the relatively 
           heavy squark mass), then the relic density of gluino will be mainly from 
           the neutralino decay. In such a case, a much stronger upper bound of about 2.2 TeV
           on the LSP mass obtained in \cite{profumo} should be applicable in order not to 
           overclose the universe (note that in our case the upper bound of 2.2 TeV is for
           gravitino LSP mass and the upper bound on the neutralino mass can
           be relaxed since now the relic density of dark matter is given by
           $\Omega_{\tilde \chi^0_1} \frac{m_{\tilde G}}{m_{\tilde \chi^0_1}}$).  

\begin{figure}[tb] 
\hspace*{-0.7cm}\epsfig{file=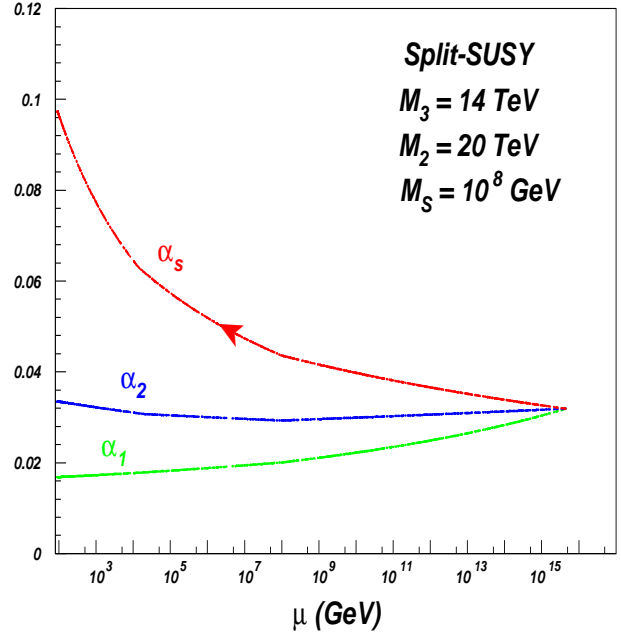,width=9.0cm, height=8.7cm}
\caption{ The one-loop running of gauge couplings in split-SUSY with fixed values of $M_3$,
          $M_2$ and sfermion mass $M_S$.}
\end{figure}

           Since our results are valid only in case that the neutralino decay to gluino 
           happens before the freeze-out of 
           gluino, we now examine the condition of this
           scenario. 
           When the gluino is as heavy as 14 TeV, its freeze-out temperature is found to 
           be about $m_{\tilde{g}}/30$ and the freeze-out time is thus about $10^{-9} sec$.
           The neutralino decays into gluino via exchanging a squark and its lifetime 
           is sensitive to the forth power of squark mass:
   \begin{equation}
   \tau_{\tilde \chi^0_1}=3\times 10^{-2} sec {\left(\frac{M_S}{10^9 GeV}\right)}^4
                    {\left( \frac{1 TeV}{ m_{\tilde \chi^0_1}}\right)}^5 ,
   \end{equation}
   where $M_S$ is squark mass. For a neutralino at the order of 10 TeV,  
   in order to let its lifetime shorter than gluino freeze-out time ($\sim 10^{-9} sec$), 
   the squark mass can be chosen to be $M_S \sim 10^8$ GeV.
  
(3) Our analyses showed that the gluino (and all other gauginos or Higgsinos) 
           must be heavier than about 14 TeV in order to provide the full dark matter 
           abundance in the scenario we considered. It is interesting to check whether or
           not such a scenario is consistent with the gauge couplings unification at some
           high energy scale. In Fig. 3 we show the one-loop running of three gauge couplings 
           for $M_3=14$ TeV, 
           $M_2=20$ TeV and squark mass $M_S= 10^8$ GeV. 
           Here,  $M_3$ is gluino mass and, just like in ref.\cite{senatore},  
           we assumed that Bino, Wino and Higgsino are all degenerate at the scale $M_2$.
           From Fig. 3 we see that starting from $M_Z$ scale
           \footnote{The starting values of  $\alpha_1$ and $\alpha_2$ at  $M_Z$ scale is 
           fixed by $\alpha^{-1}(M_Z)=128.936\pm 0.0049$ and 
           $\sin^2\theta_W(M_Z)=0.23150\pm 0.00016$\cite{data}.},  
           $\alpha_1$ and $\alpha_2$ run up to higher energy scale 
           and finally meet at a crosspoint at $\sim 10^{16}$ GeV. From this crosspoint 
           $\alpha_s$ runs back to $M_Z$ scale and ends at  $\alpha_s(M_Z)=0.098$. This 
           value is welcome since, as pointed in \cite{split}, the two-loop effects will
           enhance $\alpha_s$ at $M_Z$ scale by about 0.022. Taking into such effects,
           $\alpha_s(M_Z)$ in our scenario is just within the $2-\sigma$ range 
            $0.119\pm 0.003$ \cite{data} allowed by experiments.             
     
(4) Since in our scenario the gluino is the NLSP and all other gauginos are heavier
           than gluino, which is phenomenologically viable so far, the gauginos spectrum 
           is different from that predicted by some 
           theoretically favored models like mSUGRA. In the popular mSUGRA models, 
           for example, the colored gluino is predicted to be heavier than other gauginos 
           at the weak scale. However, such fancy models may not be chosen by nature and 
           phenomenologically we should not be restricted to them.

(5) If this dark matter scenario (LSP is gravitino produced from the late decay of 
           the metastable gluino in split-SUSY) is indeed chosen by nature, then no super 
           particles of split-SUSY can be found at the LHC except a light Higgs boson whose
           mass is upper bounded by about 150 GeV \cite{split} \footnote{This bound may be
           lowered by a few tens of GeV if right-handed neutrinos are introduced with see-saw
           mechanism\cite{cao}.}.         

\section{Conclusion}
The metastable gluino in split-SUSY can freeze out in the early universe
and then decay slowly into the LSP, providing much of the cosmic dark matter content. 
If the LSP is the gravitino produced from the late decay of the metastable gluino,
we found that the dark matter consideration can constrain the parameter space of the 
gluino mass versus the gravitino mass: in order to provide the full abundance of dark 
matter, the gluino must be heavier than 14 TeV. Therefore, if nature takes this choice 
for dark matter, no gauginos or Higgsinos are accessible at the LHC. Then no super particles
of split-SUSY can be found at the LHC except a light Higgs boson. 

\section*{Acknowlegement}
We thank Prof. Bing-Lin Young, Dr. Guangping Gao, Guoli Liu and Fuqiang Xu for discussions. 
This work is supported in part by National Natural Science Foundation of China.

\end{document}